\documentstyle[preprint,aps,epsfig]{revtex}

\def\ybco{YBa$_2$Cu$_{3}$O$_{7-\delta}$}
 
\def\xab{\xi_{ab}}

\draft

\begin{document}
%@twocolumnfalse\endcsname

\title{       Simple Model for the Variation of Superfluid Density with
                Zn Concentration in \ybco} 
\author{Jeng-Da Chai, Sergey V. Barabash, and 
D. Stroud\footnote{Corresponding author: tel.
(614)292-8140; fax (614)292-7557; email stroud@mps.ohio-state.edu}}
\address{
Department of Physics,
The Ohio State University, Columbus, Ohio 43210}

\date{\today}

\maketitle

\tightenlines

\begin{abstract}

We describe a simple model for calculating the zero-temperature superfluid
density of Zn-doped  
YBa$_2$Cu$_3$O$_{7-\delta}$ as a function of the fraction $x$ of in-plane
Cu atoms which are replaced by Zn.
The basis of the calculation is a ``Swiss cheese'' picture of a 
single CuO$_2$ layer, in which a substitutional Zn impurity creates a normal
region of area $\pi\xi_{ab}^2$ around it as originally suggested by
Nachumi {\it et al}.  Here $\xi_{ab}$ is the zero-temperature
in-plane coherence length at $x = 0$.  
We use this picture to calculate the variation
of the in-plane superfluid density with $x$
at temperature $T = 0$, using both a numerical approach 
and an analytical approximation.  
For $\delta = 0.37$, if we use the value $\xab = 18.3$ $\AA$,
we find that the in-plane superfluid decreases with
increasing $x$ and vanishes near $x_c = 0.01$ in the analytical 
approximation, and near $x_c = 0.014$ in the numerical approach.  
$x_c$ is quite sensitive to $\xab$, whose value is not widely agreed upon.
The model also predicts a peak in the real part of the 
conductivity, Re$\sigma_e(\omega, x)$, at concentrations $x \sim x_c$, and 
low frequencies, and a variation of critical current density with $x$ of
the form $J_c(x) \propto n_{S,e}(x)^{7/4}$ near percolation, where
$n_{S,e}(x)$ is the in-plane superfluid density.

%\pacs{PACS numbers: 74.60.Ge, 72.30.+q, 74.25.Nf.  KEYWORDS: Zn-doped,
%superfluid density, YBCO, percolation}

\end{abstract}
\vskip2pc 
%\narrowtext

% \begin{multicols}{2}

\section{INTRODUCTION}

In typical s-wave superconductors, nonmagnetic impurities
are not pair-breaking, and therefore have little effect on either the 
transition temperature or the superfluid density\cite{tinkham}.  
By contrast, the substitution of a nonmagnetic
impurity into a d-wave superconductor is expected to be pair-breaking, and,
therefore, should have a significant effect on both.  Since the CuO$_2$-based
high-temperature superconductors are believed to possess an order parameter
with $d_{x^2-y^2}$ symmetry\cite{dwave}, such an effect should be observable
in these materials.  Indeed, the substitution of a nonmagnetic impurity 
such as Zn (or even an insulating impurity such as He, introduced by
ion implantation) in a high-T$_c$ material such as 
YBa$_2$Cu$_3$O$_{7-\delta}$ is now well known
to dramatically suppress the zero-temperature superfluid 
density\cite{ulm,jung,karpinska,basov,bernhard,moffat}
and also to have a significant, though lesser, effect on the critical 
temperature.  

This striking behavior has provoked numerous theoretical studies
of Zn impurities and other nonmagnetic disorder
in the cuprate superconductors.  For example, 
Annett {\it et al}\cite{annett} and Hirschfeld and 
Goldenfeld\cite{hirschfeld}, have shown that the impurity scattering
can convert the $T$-linear behavior of the low-temperature
superfluid density to a $T^2$ behavior, in agreement with experiment.  
Other workers\cite{prohammer,kim,sun} have refined the theory 
so that it now agrees quite well with experiment.  Choi\cite{choi} has 
used a t-matrix formalism to calculate the Zn impurity dependence 
of the superfluid density in YBa$_2$Cu$_3$O$_{7-\delta}$
for various temperatures.
Salluzzo {\it et al}\cite{salluzzo} used a model of dirty
d-wave superconductivity to fit the impurity-dependence of
the superfluid density in a NdBa-based cuprate superconductor.  Hettler
and Hirschfeld\cite{hettler} have shown how inhomogeneities
in a d-wave order parameter can lead to both strong suppression of superfluid 
density and enhanced microwave conductivity.   

Several authors have treated the impurity problem by solving the
Bogoliubov-de Gennes equations\cite{tinkham} for the pairing amplitude,
which is position-dependent in a dirty d-wave or s-wave superconductor 
with a short coherence length.
Ghosal {\it et al}\cite{ghosal1} used this approach to
show that disorder in a 2D s-wave superconductor reduces the 
effective  superfluid density, makes the superfluid inhomogeneous, and
even breaks up the 2D layer into superconducting islands.
More recently\cite{ghosal2}, the same authors generalized this
approach to d-wave superconductors.  Zhitomirsky and 
Walker\cite{zhitomirsky} showed how
spatial fluctuations in the energy gap can reduce T$_c$ in a d-wave
superconductor.  Haas {\it et al}\cite{haas} solved the
Bogoliubov-de Gennes equations with a momentum-dependent pairing interaction
to obtain an extended gapless region in the overdoped cuprates, which they
connected to a rapid doping-induced decrease of both $T_c$ and 
superfluid density.  Franz {\it et al}\cite{franz}, using the Bogoliubov-de
Gennes approach for a spatially varying energy gap,
found that the superfluid density 
goes rapidly to zero with increasing Zn concentration, while T$_c$
decreases much more slowly with concentration, 
in reasonable agreement with experiment.
Neto\cite{neto} has treated the Zn-doped cuprates as a collection of 
fluctuating stripes modeled as an array of coupled overdamped 
Josephson junctions pinned by impurities, and has used this model to 
estimate the doping-dependence of the transition
temperature.  Finally, Uemura\cite{uemura} has discussed the low-temperature 
superfluid density in Zn-doped cuprates from the point
of view of the ``universal correlation'' between 
$T_c$ and zero-temperature superfluid density
in the underdoped cuprates.  As noted by Emery and Kivelson\cite{emery},
such a connection would exist if $T_c$ were controlled by phase 
fluctuations rather than by the vanishing of the gap amplitude.

Most of these models require a rather elaborate calculation to find
the influence of non-magnetic impurities on the superfluid density.  By
contrast, the goal of this paper is to analyze the behavior to be expected of
a very simple model of the Zn-doped cuprate superconductors which can
be worked out by intuitively appealing calculations. 
Although this model is difficult to derive from first principles, it
may still be of value, because it can be used to make
simple predictions for specific experiments, and to understand
trends in those experiments.

The model we study was first described by Nachumi 
{\it et al}\cite{nachumi}, and is sometimes known 
as the ``Swiss cheese'' model.   In this model, the Zn
atoms are assumed to replace the Cu atoms substitutionally within the
CuO$_2$ layer, which is believed to be the locus of superconductivity in
the cuprate superconductors, and to disrupt the $d_{x^2-y^2}$ order parameter,
producing a region approximately of radius $\xi_{ab}$ within which the
layer is normal.  (Here $\xi_{ab}$ is the zero-temperature in-plane coherence
length of YBa$_2$Cu$_3$O$_{7-\delta}$.)
This disruption is plausible, because $\xi_{ab}$ is very
small in the cuprate superconductors (probably on the order of 10 - 20 $\AA$
at low temperatures, as discussed further below).  Indeed,  
some dramatic experimental evidence supporting both the 
Swiss cheese model and the d-wave order parameter picture 
has recently been presented by Pan {\it et al}\cite{pan}.  These workers 
showed, using scanning tunneling microscopy,
that the gap was significantly suppressed near the Zn atoms, and that the
density of states of single-particle excitations near the Zn atoms exhibited
a fourfold symmetry.  

In the present paper, we extend the approach of Nachumi {\it et al} by 
analyzing the Swiss cheese model within a simple percolation picture.  
Our approach differs from that of Nachumi {\it et al} by accounting for
the connectivity of the superconducting part of the CuO$_2$ layer.  By
contrast, in their approach, the 
superfluid density is directly proportional to the
superconducting area.  The consequences of our approach are 
easily worked out if we use percolation theory and
interpret the CuO$_2$ layer as
an inhomogeneous superconductor with spatially distinct superconducting
and  normal region.
The resulting variation of superfluid density with Zn concentration
seems to agree somewhat better with experiment than does the simplest
version of the Swiss-cheese model.  

The remainder of this paper is organized as follows.  In the next section,
we describe our percolation model in detail.  
The numerical results following from this model
are presented in Section III.  Finally, in Section IV, we discuss our numerical
results, briefly compare them to experiment, and analyze the limitations and
predictions of the model.

\section{MODEL}

\subsection{Geometrical Assumptions}

Fig.\ 1 shows our picture of a single CuO$_2$ layer in Zn-doped
YBa$_2$Cu$_3$O$_{7-\delta}$, in which a fraction $x$ of the Cu atoms are 
replaced at random by Zn.  Within the 
``Swiss cheese'' model\cite{nachumi}, 
each Zn atom completely suppresses superconductivity
in a region of area $\pi\xab^2$  around it.  Thus, as $x$ increases, a
%%Note: THIS IS THE FIRST DEFINITION OF THE SWISS CHEESE PICTURE
smaller and smaller areal fraction of the CuO$_2$ plane is superconducting.
Eventually, for a large enough $x$, the remaining superconducting region
is disconnected and the effective superfluid density must vanish.

There are two parameters in our geometrical model: the fraction $x$ of in-plane
Cu atoms which are replaced by Zn, and the quantity $\xi_{ab}/a$,
which is the ratio of the zero-temperature in-plane coherence length 
$\xi_{ab}$ (evaluated at $x = 0$) to the  Cu-Cu distance $a$.  Neither
parameter is trivial to determine experimentally.   For example, $x$ is not
necessarily equal to $c$, the {\em overall} fraction of Cu atoms which are
replaced by Zn.  In fact, it is believed that $x \sim 3c/2$, i.\ e.,
that all the Zn dopant atoms go into the CuO$_2$ planes \cite{gupta}. 
Obviously, the connection between $x$ and $c$ is important in making quantitative
comparisons between theory and experiment.   Also,
the magnitude of $\xab$ is not universally agreed upon for any value
of $\delta$.  Nachumi {\it et al}\cite{nachumi}, for example, quote 
$\xab = 18.3\AA$ at $c = 0$ and $\delta = 0.37$, 
extrapolating from $c = 0.01$.  
Since $a = 3.86 \AA$ for this value of $\delta$\cite{kakurai},
this gives $\xab/a = 4.74$.  
We shall later discuss the effects of assumptions about $\xab$ on the
predictions obtained from this model.

\subsection{Assumptions about $\sigma_S$ and $\sigma_N$}.

To estimate the superfluid density as a function
of $x$, we assume that the 
region more than a distance $\xi_{ab}$ from any Zn atom has a complex
conductivity $\sigma_S(\omega)$ characteristic of a superconductor.  
We obtain this conductivity by using the first London equation, 
as applied to a CuO$_2$ plane at $x = 0$:
\begin{equation}
\Lambda\frac{\partial {\bf J}_s}{\partial t} = {\bf E},
\end{equation}
where ${\bf J}_s$ is the supercurrent density and ${\bf E}$ is the electric
field.  In gaussian units, 
\begin{equation}
\Lambda = \frac{4\pi \lambda_{ab}^2}{c^2} = \frac{m^*}{n_Sq^2},
\end{equation}
where $\lambda_{ab}$ is the in-plane penetration depth (evaluated at $x = 0$),
and $m^* = 2m_e$ is the mass and $q = 2|e|$ the charge of a Cooper pair.  
Fourier transforming gives
${\bf J}_s(\omega) = \sigma_S(\omega){\bf E}(\omega)$, where 

\begin{equation}
\sigma_S(\omega) = \frac{iA}{\omega}
\end{equation}
with 
\begin{equation}
A = \frac{q^2n_S}{m^*} = \frac{c^2}{4\pi\lambda_{ab}^2}.
\label{eq:defa}
\end{equation}
We assume that the region {\em within} a distance
$\xi_{ab}$ of a Zn atom is 
simply a normal metal with frequency-independent conductivity $\sigma_N$.

Clearly, this simple model can only be appropriate for
$\omega \ll \omega_0$, where $\omega_0 = A/\sigma_N$.  At higher
frequencies, there will be effects arising from the superconducting
energy gap, which will produce additional 
structure in $\sigma_S(\omega)$.   Also, this inductive form also implies
no absorption at low frequencies in the limit $x \rightarrow 0$.
In reality, many of the cuprate superconductors do
exhibit low-frequency absorption, even at low
temperatures\cite{orenstein}.  This extra absorption can readily be included
within the present model by adding to $\sigma_S$ a parallel normal
conductivity.  We have carried out a few calculations including this
extra term; as discussed below, they do not affect the $x$-dependent
superfluid density.

\subsection{Definition of Effective Superfluid Density $n_{S,e}(x)$}

We may infer the effective superfluid density $n_{S,e}(x)$
from an effective London equation
\begin{equation}
\Lambda_e\frac{\partial\langle {\bf J}\rangle}{\partial t} = 
\langle {\bf E}\rangle
\label{eq:london1}
\end{equation}
which relates the space-averaged current density $\langle{\bf J}\rangle$ 
to the space-averaged electric field $\langle {\bf E}\rangle$.  $\Lambda_e(x)$
is related to $n_{S,e}(x)$ through a
concentration-dependent
generalization of eq.\ (2):
\begin{equation}
\Lambda_e(x) = \frac{4\pi}{\lambda_{ab;e}^2(x)}{c^2} = \frac{m^*}{n_{S,e}(x)q^2}.
\label{eq:london2}
\end{equation}
In the frequency domain, eqs.\ \ref{eq:london1} and \ref{eq:london2}
may be written
\begin{equation}
\langle {\bf J}(\omega)\rangle = i\frac{n_{S,e} q^2}{m^*\omega} 
\langle{\bf E}(\omega)\rangle \equiv
\sigma_e(\omega)\langle{\bf E}(\omega)\rangle,
\end{equation}
where $\sigma_e(\omega, x)$ is the effective complex conductivity.
Therefore,
\begin{equation}
n_{S,e}(x) = \frac{m^*}{q^2}Lim_{\omega \rightarrow 0}\omega 
Im \sigma_e(\omega,x),
\label{eq:super}
\end{equation}
and, in order to compute $n_{S,e}(x)$, we need to calculate 
$\sigma_e(\omega,x)$ at sufficiently low frequencies.

The ratio $n_{S,e}(x)/n_{S,e}(0)$ can also be calculated more directly by using
a homogeneity property of $\sigma_e$\cite{bergman}.  In the present context, 
this property states that
\begin{equation}
\sigma_e(\sigma_S, \sigma_N, x) = \mu\sigma_e(\sigma_S/\mu,
\sigma_N/\mu,x)
\end{equation}
Here $\sigma_e(\sigma_S, \sigma_N, x)$ is the effective complex conductivity
of a two-component system made up of constituents with conductivities
$\sigma_S$ and $\sigma_N$ and in-plane Zn concentration $x$ 
{\em in a particular geometry}, and $\mu$ is
an arbitrary constant.  In effect, homogeneity means that (for fixed geometry
and fixed $x$)
if the conductivities of each constituent are multiplied by
a certain factor $1/\mu$, the {\em effective} conductivity is multiplied
by that same factor $1/\mu$.  In particular, if $\mu = \sigma_S$, we
get
\begin{equation}
\sigma_e(\sigma_S, \sigma_N, x) = \sigma_S\sigma_e(1, \sigma_N/\sigma_S, x).
\label{eq:homog}
\end{equation} 
Hence, for any $x$, 
\begin{eqnarray}
\frac{n_{S,e}(x)}{n_{S,e}(0)} & = &\frac{Lim_{\omega\rightarrow 0}\omega
Im\sigma_e(\sigma_S(\omega), \sigma_N, x)}{Lim_{\omega\rightarrow 0}\omega
Im\sigma_e(\sigma_S(\omega), \sigma_N, 0)} \\
& = &\frac{\sigma_e(1, 0, x)}{\sigma_e(1, 0, 0)}.
\label{eq:ratio}
\end{eqnarray}
Here we have used the fact that $n_{S,e}(0)$, the superfluid density at $x = 0$,
is given by $n_{S,e} = m^*A/q^2$, and also the fact that
Lim$_{\omega \rightarrow 0}\sigma_N/\sigma_S(\omega) = 0$.

\subsection{Two Methods for Calculating $n_{S,e}(x)$}

We have evaluated $n_{S,e}(x)$ [or equivalently, $\sigma_e(\omega,x)$] 
by two
complementary methods.  The first method is a finite-element
technique, and should become very accurate for a sufficiently large numerical
sample, given the geometrical assumptions of the model.  We consider an 
L $\times$ L region of a CuO$_2$ plane,
assuming that the Cu ions sit on a square lattice of lattice constant $a$, and
choosing $L = N_xa$ where $N_x$ is an integer.  Then, using a random number 
generator, we replace a fraction $x$ of the Cu ions
by Zn.  We then assign conductances $\sigma_S(\omega)$ or $\sigma_N$ to each
bond connecting nearest neighbor Cu sites, by the following rule.  If the
entire length of the bond is at least a distance $\xab$ from any Zn site, it
is assigned a conductance $\sigma_S(\omega)$; otherwise, it is assigned
$\sigma_N$.   Next, we calculate $\sigma_e(\omega,x)$ using a standard technique
known as the Y-$\Delta$ transformation\cite{frank}, which is very efficient
and accurate for two-dimensional impedance networks.  This technique is
approximate only insofar as it is limited to finite networks.  We call
this method the random conductance approach (RCA).

Our second method for calculating $\sigma_e(\omega,x)$ is a
simple analytical approximation, the effective medium approximation
(EMA)\cite{bruggeman,landauer,bergman}.  For the assumed geometry, this
approximation is implemented as follows.   First, we calculate $p$, the
areal fraction occupied by the superconducting material, as a function of
$x$, using the relation
\begin{equation}
p = Lim_{S \rightarrow \infty}\left(1 - \frac{\pi \xab^2}{S}\right)^{N_{Zn}},
\end{equation}
where $N_{Zn}$ is the number of Zn ions in a plane of area $S$.  This limit
is readily evaluated using the relation $x=N_{Zn}a^2/S$, with the result
\begin{equation}
p = \exp\left(-x\frac{\pi\xab^2}{a^2}\right).
\end{equation}
Given $p$, one can calculate
$\sigma_e$ using the standard EMA relation for this two-dimensional
system, namely,
\begin{equation}
p\frac{\sigma_N-\sigma_e}{\sigma_N+\sigma_e} + (1-p)\frac{\sigma_S-\sigma_e}
{\sigma_S+\sigma_e} = 0.
\end{equation}
The physically correct solution to this quadratic equation 
is the one which satisfies
Im$\sigma_e > 0$.  Of course, ideally, the EMA is suited for application to
a geometry in which the $N$ and $S$ components are distributed in a
symmetrical fashion, whereas here, because of the 
assumption of excluded area, the $N$ and $S$ are not distributed
symmetrically.  Nonetheless, it is a reasonable approximation even for
this nonsymmetric geometry.

Once $\sigma_e(\omega)$ has been calculated at a 
given $x$ by either of the methods described above, 
the effective superfluid density can be inferred 
from the relation (\ref{eq:super}).
We can also compute 
the ratio $n_{S,e}(x)/n_{S,e}(0)$ directly, 
using eq.\ (\ref{eq:ratio}).  That is, for a given $x$, we decide which are 
the superconducting bonds, using the method described earlier in this section,
and assign these bonds conductances of unity.  
The remaining normal bonds are assigned conductances
of zero.  The effective conductance is then evaluated using either the
EMA or the RCA, and the ratio $n_{S,e}(x)/n_{S,e}(0)$ is then obtained
using eq.\ (\ref{eq:ratio}).

\section{RESULTS}

In carrying out the RCA calculations, we have generally used a
square networks of size ranging from $160 \times 160$ to $500 \times 500$ Cu
sites.  In order to reduce statistical 
fluctuations, the RCA results quoted below are averaged over a number of
different realizations of the disorder for each $\omega$ and $x$.  
These realizations are ``correlated'' in the sense that, for each realization,
we increase $x$ by adding a few impurities to the configuration of the
previous $x$ value.  In evaluating the EMA conductivities, no such 
averaging is necessary, since the approximation can be evaluated 
analytically in the limit of a very large sample.

Our results for $n_{S,e}(x)/n_{S,e}(0)$ 
are shown in Fig.\ 2 for several choices of the parameter 
$\xi_{ab}/a$.  The full, dashed, dot-dashed, and dotted curves 
are the EMA results. 
The diamonds are the results of the RCA calculations at $\xi_{ab}/a =
4.74$, using eq.\ (12), for two correlated realizations of a $500 \times 500$
lattice.  There are several striking features.   
First of all, in both methods
$n_{S,e}(x)/n_{S,e}(0)$ monotonically decreases with increasing $x$, 
reaching zero at a percolation threshold $x_c$ which depends on 
$\xi_{ab}/a$.  Moreover,
$x_c$ increases with decreasing $\xi_{ab}/a$.  This relationship is not
surprising: a larger $\xi_{ab}/a$ implies that a larger area 
is converted from superconducting to normal by a single Zn 
impurity, within this model.    Also, the 
EMA and RCA agree quite well over most of the concentration
range, except near the percolation threshold.  The EMA percolation 
threshold always corresponds to $p = 0.5$\cite{bergman}.  
The RCA threshold is higher, because the superconducting
fraction remains connected to a somewhat higher fraction of Zn than is
predicted by the EMA.   

In Figs.\ 3 and 4, we show Im$\sigma_e(\omega, x)/\sigma_N$ and 
Re$\sigma_e(\omega, x)/\sigma_N$
as a function of x at $\omega = 0.001\omega_0$, where 
$\omega_0 = A/\sigma_N$ and $A$
is defined in eq.\ (\ref{eq:defa}).  We use using parameters thought to be 
appropriate for the case $\delta = 0.37$, namely, $\xi_{ab}/a =
4.74$\cite{kakurai}.   In both Figures, the dashed line represents the
EMA; the diamonds denote the RCA (average over 20 correlated realizations
of a $160 \times 160$ lattice) and the the open squares are the RCA, averaged
over 2 realizations of a $500 \times 500$ lattice.  Fig.\ 3 shows that
Im$\sigma_e(\omega, x)/\sigma_N$ closely mirrors the behavior of 
$n_{S,e}(x)/n_{S,e}(0)$
shown in Fig.\ 2, as expected from the homogeneity relations (11) and (12).
The
apparent percolation threshold is again somewhat larger ($x_c \sim 0.014$) in 
the RCA
than in the EMA ($x_c \sim 0.01$).  For values of $x$ well below $x_c$,
Im$\sigma_e(\omega = 0.001\omega_0, x)/\sigma_N$, 
like $n_{S,e}(x)/n_{S,e}(0)$, decreases
linearly with $x$.  This linear behavior is well known in the
EMA\cite{bergman}.  

Fig.\ 4 shows that Re$\sigma_e(\omega,x)$ has a strong 
peak occurring near the percolation threshold 
($x_c \sim 0.01$ in the EMA and $\sim 0.014$ within the RCA for our choice
of $\xab/a$).   The physical origin of this peak is discussed in the next
section.  We believe that the greater half-width of the RCA results 
arises because the calculation is carried out for a finite sample,
and also because the average is taken over a number of different
realizations, each of which has a slightly different percolation threshold
for these finite samples.  In support of this picture, note that the RCA
half-width is smaller for the larger ($500 \times 500$)
samples, for which these fluctuations should be smaller.   The slight
random fluctuations in these curves as a function of $x$ are, we believe,
also due to these finite-size effects.  By contrast, the EMA
results are obtained analytically for an effectively infinite sample; so these
finite-size fluctuations play no role.  Note that, even for a $500 \times
500$ sample, there are still only 375 impurities at $x = 0.015$; so these
fluctuations are still significant in the RCA.

In Fig.\ 5, we show Re$\sigma_e(\omega, x)/\sigma_N$ for 
$\omega = 0.0001\omega_0$, once
again using $\xab/a = 4.74$, as calculated in the EMA and in the 
RCA for two different sample sizes.  The EMA peak is clearly stronger 
for $\omega = 0.0001\omega_0$ than $\omega = 0.001\omega_0$.  The RCA peak is also stronger,
especially for the larger sample where the peak height is not washed out by
finite size effects.  The apparent double peak for the $500 \times 500$ sample
is, we believe, just the result of finite-size fluctuations, which are
magnified at such low frequencies where the conductivity contrast 
$|\sigma_S/\sigma_N|$ is larger.      

Finally, we have calculated $\sigma_e(\omega, x)/\sigma_N$ within the EMA
for the same model as before, again with $\xab/a = 4.74$, 
but a slightly different choice for
$\sigma_S(\omega)$.  Namely, we write
\begin{equation}
\sigma_S(\omega) = \frac{iA}{\omega} + \sigma_N^\prime,
\label{eq:sigs1}
\end{equation}
where we arbitrarily choose $\sigma_N^\prime = \sigma_N$.  The motivation
behind this choice is that, in a material such as \ybco, which is
believed to have an order parameter with $d_{x^2-y^2}$ symmetry, the in-plane
conductivity is expected to have a quasiparticle contribution, even
at very low temperatures.  The additional term $\sigma_N^\prime$ is a 
crude way of modeling this contribution.  
The resulting Re$\sigma_e(\omega,x)$ is plotted versus $x$ at $\omega =
0.001\omega_0$, both for the model (\ref{eq:sigs1}) and for a purely inductive
$\sigma_S = iA/\omega$.   Clearly, the peak near $x_c$ is little affected by the change, 
but there is now, in addition, some extra contribution to 
Re$\sigma_e(\omega,x)$ for $x < x_c$.  Although this extra contribution
to $\sigma_S$ does slightly change Re$\sigma_e(\omega,x)$ at 
finite $\omega$, the homogeneity 
relations (11) and (12) imply that it has no effect
on $n_{S,e}(x)$, either in the EMA or in the RCA.

\section{DISCUSSION}

It is of interest to compare our calculated values of $x_c$ 
with the experimentally observed critical concentration
for YBa$_2$(Cu$_{1-c}$Zn$_c$)$_3$O$_{7-\delta}$ at $\delta = 0.37$.  
The critical value of $c$ is between $0.02$ and 
$0.03$\cite{bernhard,haran,xiang,jayaram,mizuhashi,raffo}
(probably closer to $0.03$, judging from
in-plane resistivity measurements\cite{mizuhashi}), as
compared to our calculated critical value $x_c \sim 0.014$ for this 
$\delta$ [cf.\ Fig.\ 2].  But a direct comparison of these values is difficult:
the in-plane Zn concentration may differ from $c$.  If in fact
$x \sim 3c/2$, 
%%NEED REFERENCE HERE
as mentioned above, then our model seems to underestimate
$x_c$; the underestimate is less
in the RCA than in the more approximate EMA.
But this apparent discrepancy in $x_c$ may not be very significant, for
two reasons.  First, $x_c$ is quite sensitive to the parameter $\xab/a$: 
a change of $\xab$ only from $4.74$ to $3$ increases $x_c$ by more than a 
factor of 2 in the EMA, and presumably by a similar amount in the RCA.  As
already mentioned, this parameter is not known experimentally with great
precision.  Secondly, our model assumes that the CuO$_2$ layer 
undergoes an {\em discontinuous} change
from S to N character at a distance $\xab$ from a Zn impurity.   It would be
more realistic to assume a gradual change.   Although such a transition 
region might be difficult to include in the present model, its qualitative
effect would probably be to reduce the effective radius of the normal region
to a value smaller than $\xab$.  Such a reduction would, according to our results,
{\em increase} $x_c$, making it closer to the reported 
value\cite{mizuhashi,raffo}.

It is of interest to compare the present model to several others in the
literature.  In the original model of Nachumi {\it et al}\cite{nachumi}, 
the superfluid density is proportional to the superconducting areal fraction, 
i.\ e., to the fraction of the CuO$_2$ plane which is more than a 
distance $\xi_{ab}$ from any
Zn impurity.  The present model improves on the implementation of this Swiss
cheese model by taking into account effects of connectivity of the
superconducting parts.  This treatment is consistent with the generalized
London equation [eq.\ (5)].  In another class of 
models (see, e.\ g., Refs.\ \cite{ghosal1,franz}) 
$n_{S,e}(x)$ is calculated using the Bogoliubov-de
Gennes equation with some form of static disorder.  This model, like the
Swiss-cheese model, also leads to a spatially varying gap function or pairing
potential.  Although this model seems quite different from ours, we speculate
that both actually have much of the same physical content. 
The difference is that in the models of Refs.\ \cite{ghosal1,franz}, the
disorder is on an atomic scale and treated microscopically, whereas in
the present case the disorder is on a greater length scale and treated
within continuum electrodynamics.   Nonetheless, both lead to surprisingly
similar behavior (superfluid density diminishing rapidly with increasing
$x$ and vanishing at a critical concentration).   We do not know how
to connect the two on a more rigorous basis, however.

A nontrivial prediction of the present model 
is the existence of a peak in Re$\sigma_e(\omega)$
near $x_c$.  If such a peak were observed in experiments, it could be viewed as
strong evidence in favor of the present model.  
This ``percolation peak'' is reminiscent of the well-known 
fluctuation peak in Re$\sigma(\omega)$ near T$_c$ in a homogeneous 
superconductor, but there are significant differences in the underlying
physics.  The fluctuation peak results from slow motions of 
superconducting fluctuations with 
size of order $\xab(T)$.  The spatial extent of these
fluctuations diverges near $T_c$, as does the time scale on which they
move; this diverging time scale leads to a strong low-frequency peak in
the conductivity.  By contrast the percolation peak involves 
motion of charge carriers through a {\em static} disordered structure; the
frequency-dependence arises from the slow motion of these charges through
the weakly connected structure (but {\em static} structure
of superconductor that exists, in our model,
near the $p_c$ (or $x_c$).  The characteristic length for this structure is
the so-called percolation correlation length $\xi_p$, which diverges near
$p_c$ on both sides of the percolation transition\cite{bergman}.  

The existence of this percolation peak can be understood from the 
following crude argument.  On the normal metal side of the percolation
threshold, the static d.\ c.\ conductivity is finite, 
but should diverge as $p \rightarrow p_c$.  This divergence is due to the
ever-larger superconducting regions, which come closer and closer to shorting
out the voltage across the normal metal as $p_c$ is approached.  This
divergence should disappear as the frequency is increased, because the
contrast in conductivities between superconducting and normal regions will
become much smaller in absolute value (that is, the ratio $|\sigma_N/\sigma_S|$
should approach unity).  Therefore, at a fixed $p$ just above $p_c$ (that is,
on the normal metal side), Re$\sigma_e(\omega,
p)$, should show a strong maximum at $\omega = 0$, and this maximum should
diverge as $p \rightarrow p_c$.  
Since $\sigma_e(\omega, p)$ should vary continuously across the 
percolation threshold, except at exactly $\omega = 0$, we expect the same 
behavior on the superconducting side of the
threshold.  This is qualitatively the behavior we see in both our EMA and
RCA results.

The behavior of this ``critical peak'' can be estimated from the standard
scaling theory of the percolation threshold (see, for example, 
Ref.\ \cite{bergman}).  This scaling approach is based on the assumption
that the percolation threshold is a critical point described by a single
diverging length $\xi_p$;
if that assumption
is correct, then the scaling approach is believed to be exact.  As applied
to the present model, and using the homogeneity relations (11) and (12).
this scaling approach dictates that $\sigma_e$ have the form
\begin{equation}
\sigma_e(1, \sigma_N/\sigma_S, x) = \sigma_S|\Delta
x|^tF\left(\frac{\sigma_N/\sigma_S}{|\Delta x|^{s+t}}\right),
\end{equation}
where $\Delta x = x - x_c$, and $s$ and $t$ are standard percolation exponents.
The function $F(z)$ of the complex variable $z$ has the limiting behaviors
\cite{bergman}
\begin{equation}
F(z)   \sim    C_1z, |z| \ll 1; \Delta x > 0; 
\end{equation}
\begin{equation}
F(z)  \sim  C_2,  |z| \ll 1; \Delta x < 0; 
\end{equation}
\begin{equation}
F(z)  \sim  C_3z^{t/(t+s)},  |z| \gg 1,  
\end{equation}
where $C_1$, $C_2$, and $C_3$ are constants.   This form implies that
precisely at $x = x_c$ both the real and imaginary parts of $\sigma_e$
satisfy
\begin{equation}
Re\sigma_e(\omega, x_c), Im\sigma_e(\omega, x_c) 
\propto \sigma_S^{s/(t+s)}\sigma_N^{t/(t+s)}
\propto \omega^{-s/(t+s)}.
\end{equation}

In conventional 2D percolation theory,
$s = t \sim 1.30$; so both Re$\sigma_e(\omega, x_c)$ and
Im$\sigma_e(\omega, x_c)$ would be proportional to $\omega^{-1/2}$.
The present ``Swiss cheese'' model may not, however, be in the same
universality class as percolation on a lattice. Instead, it may fall into
the class of certain continuum percolation models\cite{halperin}.  In
the present case, the exponent $t$ and $s$ should have the values 
corresponding to another ``Swiss cheese'' model in 2D, 
in which the area between the holes is conducting while the holes are 
insulating.  According to the discussion of Ref.\ \cite{halperin}, 
the Swiss cheese $t$ and $s$ are, in fact, unchanged from their lattice 
values of 1.3 in two dimensions.  
Therefore, one still expects that both Re$\sigma_e(\omega, x_c)$ and
Im $\sigma_e(\omega, x_c)$ should have a 
$\omega^{-1/2}$ frequency dependence in this model.

It is also of interest to consider the predictions of the present model
regarding the {\em critical current density} $J_c(x)$ in Zn-doped \ybco.
In this case, we use the results of continuum percolation 
theory as applied to $J_c(x)$\cite{lobb}, or equivalently, to $J_c(p)$.
In two dimensions, for the present Swiss-cheese
model, it is found\cite{lobb} that $J_c(p) \propto (p - p_c)^v$.  The critical
exponent $v$ is estimated as $v = (\nu + 1)(d -1)$ for Swiss cheese in 
$d$ dimensions, where $\nu$ is the percolation 
correlation length exponent defined by $\xi_p \propto |p-p_c|^{-\nu}$ .  
Thus, in two dimensions
$v = \nu + 1$.   The conductivity exponent $t$ is unchanged from its lattice
value in $d = 2$.  Using the approximations $t \sim \nu =4/3$ in $d =
2$\cite{lobb1},
we obtain $v \sim 7t/4$.  Equivalently,
this may be written (for $p$ near the percolation threshold)
$J_c(p)/J_c(1) \sim [n_{S,e}(p)/n_S]^{7/4}$.  In terms of $x$, 
$J_c(x)/J_c(0) \propto [n_{S,e}(x)/n_{S,e}(0)]^{7/4}$.  In summary, $J_c(x)$ in
Zn-doped YBa$_2$Cu$_3$O$_{7-\delta}$ is predicted to go to zero near the 
percolation threshold even more rapidly than does $n_{S,e}(x)$.  
It would be of interest if this prediction could be tested experimentally. 

The present model does not give any information directly 
about the superconducting transition temperature $T_c(x)$.  Some experiments
on Zn-doped YBa$_2$Cu$_3$O$_{7-\delta}$ indicate that $T_c(x)$ decreases much 
more slowly with $x$ than does $n_{S,e}(x, T=0)$ (see, for example, Ref.\
\cite{bernhard}).  Although we have no
quantitative model for this behavior, a simple qualitative argument does
suggest a explanation.  Let us suppose that $x < x_c$, 
so that the superconducting portion of the material
is connected throughout the plane.  
In that case, if {\em thermal fluctuation effects} can be neglected, then
$T_c(x)$ (as measured by a vanishing d.\ c.\ resistivity)
should be {\em independent} of $x$ for $x < x_c$, and zero for $x > x_c$.  In
this limit, clearly $T_c(x)/T_c(0) > n_{S,e}(x)/n_{S,e}(0)$.
In reality, thermal fluctuations should reduce $T_c(x)$ below its expected
value in the absence of fluctuations, but this inequality should still hold,
as observed experimentally.  

In summary, we have presented a simple model for the variation of superfluid
density with in-plane Zn concentration $x$ in Zn-doped \ybco.  The model
is based on the ``Swiss cheese'' picture suggested by Nachumi {\it et
al}\cite{nachumi}.  Using this model, we calculate the ratio
$n_{S,e}(x)/n_{S,e}(0)$
for Zn-doped YBa$_2$Cu$_3$O$_{7-\delta}$ as a function of the ratio 
$\xi_{ab}/a$, using two
different approximations.  The model predicts a critical Zn concentration
above which $n_{S,e}(x)$ is zero, in rough 
agreement with experiment
for Zn-doped YBa$_2$Cu$_3$O$_{7-\delta}$ at $\delta = 0.37$.  
The model also predicts a low-frequency
peak in Re$\sigma_e(\omega)$ near $x_c$, the existence of which has apparently
not been checked experimentally.  Finally, the model suggests that
the low-temperature critical current ratio 
$J_c(x)/J_c(0) \sim [n_{S,e}(x)/n_{S,e}(0)]^{7/4}$ near $x = x_c$.  
The observation of these 
features would give additional experimental support to the present simple 
model.  
                                   
\section{ACKNOWLEDGEMENTS}

      We would like to thank Professor Thomas R. Lemberger for helpful
discussions. Jeng-Da Chai thanks L. J. Schradin for valuable discussions
regarding computation and M. E. Howard for useful comments on the manuscript.
This work has been supported by NSF Grant No. DMR97-31511.

\newpage

\begin{figure}
 \epsfig{file=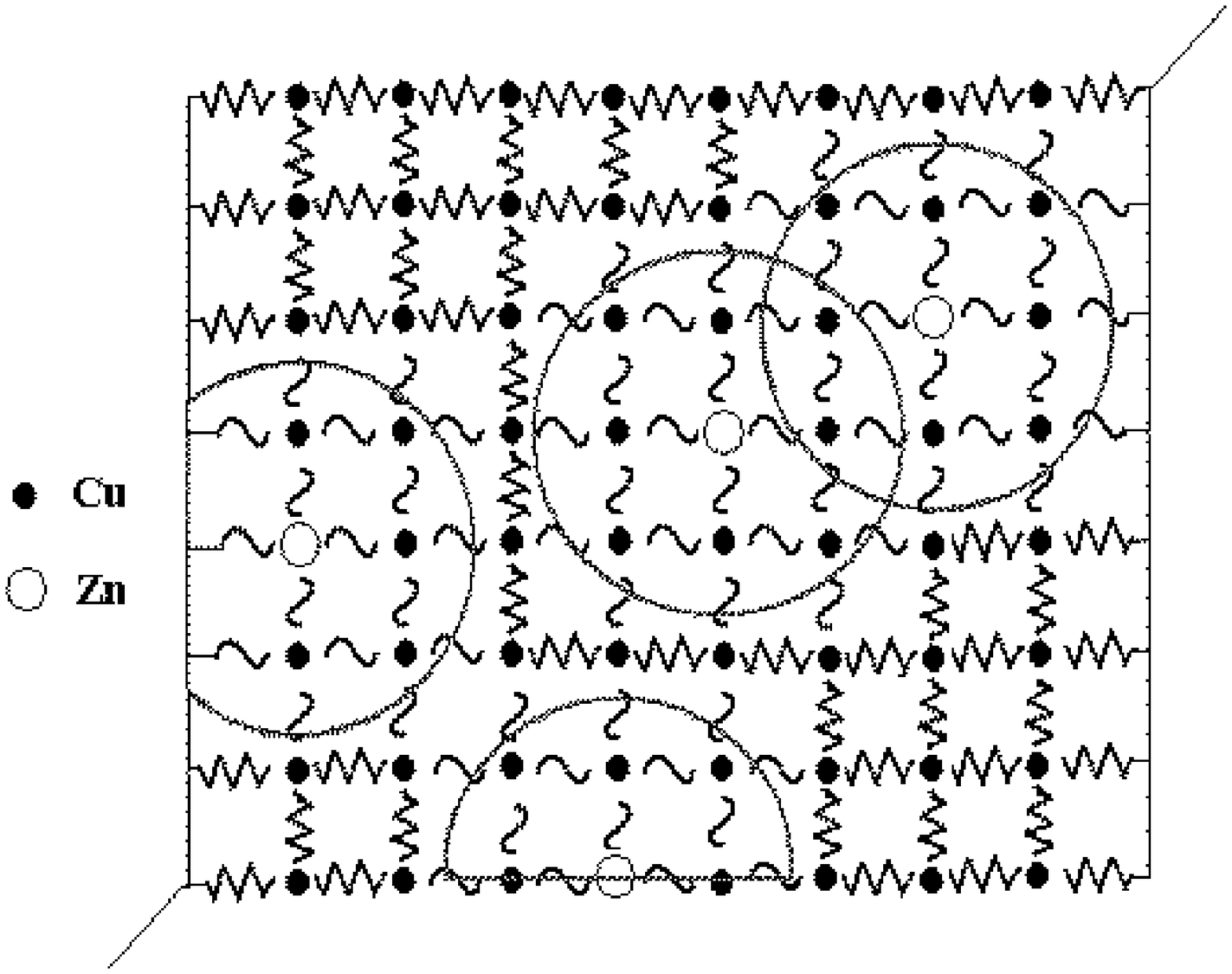, width=5in}
\label{fig1}
\caption{
Schematic diagram of model used to calculate effective superfluid
density of Zn-doped YBa$_2$Cu$_3$O$_{7-\delta}$.  
Shown is a sketch of a section of a CuO$_2$ plane.
The filled circles represent Cu ions.  The open circles represent Zn
ions, which have substitutionally replaced the Cu ions.  
$O$ ions (not shown) are situated in the middle of each bond.  The
resistor-like and wavy lines represent superconducting and normal bonds
in our model.  The large circles, of radius $\xab$, represent the 
regions which are assumed to have
been driven normal by the Zn ions.  The two diagonal lines represent the
leads used in the application of the Y-$\Delta$ transformation.
}
\end{figure}
\newpage

\begin{figure}
 \epsfig{file=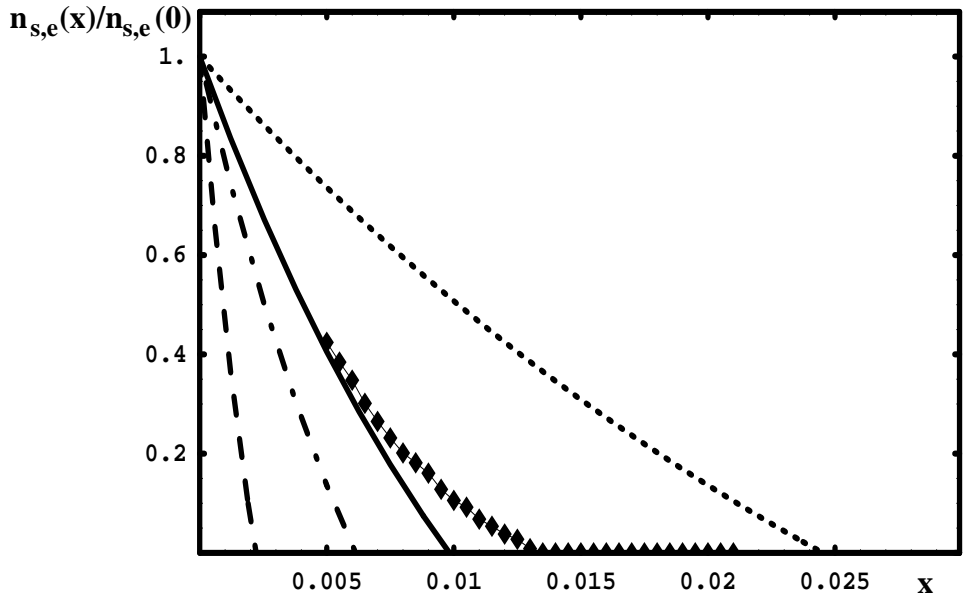, width=5in}
\vskip5.pc
\caption{
Relative zero-temperature superfluid density 
$n_{S,e}(x)/n_{S,e}(0)$, plotted as a function
of in-plane Zn atomic concentration $x$ for various values of the ratio
$\xi_{ab}/a$, where $\xi_{ab}$ is the zero-temperature 
in-plane coherence length and
$a$ the lattice constant (Cu-Cu spacing).  Dotted, solid, dot-dashed and dashed
curves correspond to EMA calculations carried out within 
the ``Swiss cheese'' model as described in the text, for
$\xi_{ab}/a = $ 3, 4.74, 6 and 10 respectively.   Filled diamond symbols
are results for $\xi_{ab}/a = 4.74$ but calculated in the RCA.  Line segments
connecting the diamonds are guides to the eye.  
Note the difference in the percolation threshold: 
$x_c$ is greater for the RCA than for the EMA.
}
\label{fig2}
\end{figure}
\newpage

\begin{figure}
 \epsfig{file=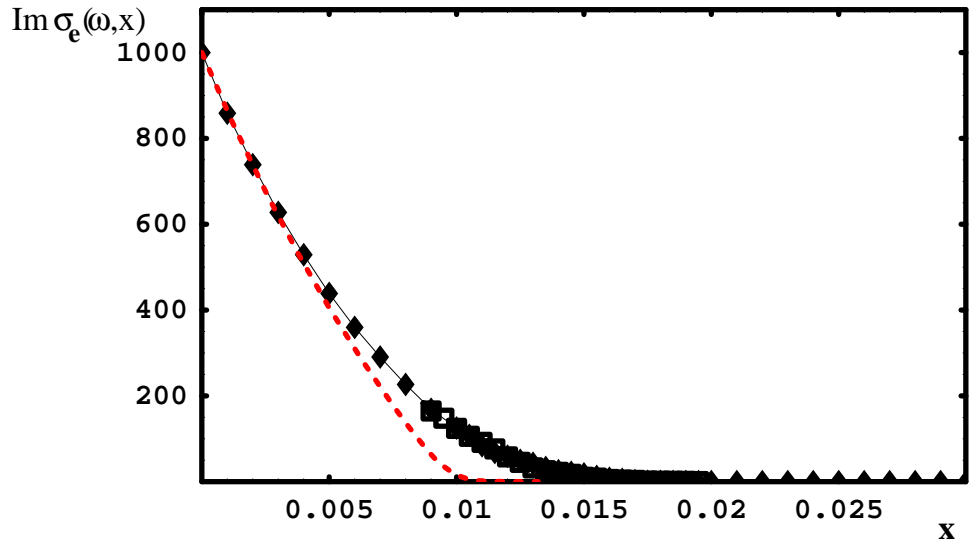, width=5in}
\vskip5.pc
\caption{
Im$\sigma_{e}(\omega, x)/\sigma_N$ versus
in-plane Zn atomic concentration $x$ for $\omega = 0.001\omega_0$, where 
$\omega_0 = A/\sigma_N$, assuming
$\xi_{ab}/a = 4.74$.  We use $\sigma_S(\omega) = iA/\omega$,
$\sigma_N = $ const.  Filled diamonds are RCA calculations
for a 160 $\times$ 160 lattice, averaged over twenty 
correlated realizations; open squares are RCA for two correlated realizations
of a $500 \times 500$ lattice).  Solid line through these points is a guide
to the eye.  Dashed line: effective-medium approximation (EMA).  
}
\label{fig3}
\end{figure}
\newpage

\begin{figure}
 \epsfig{file=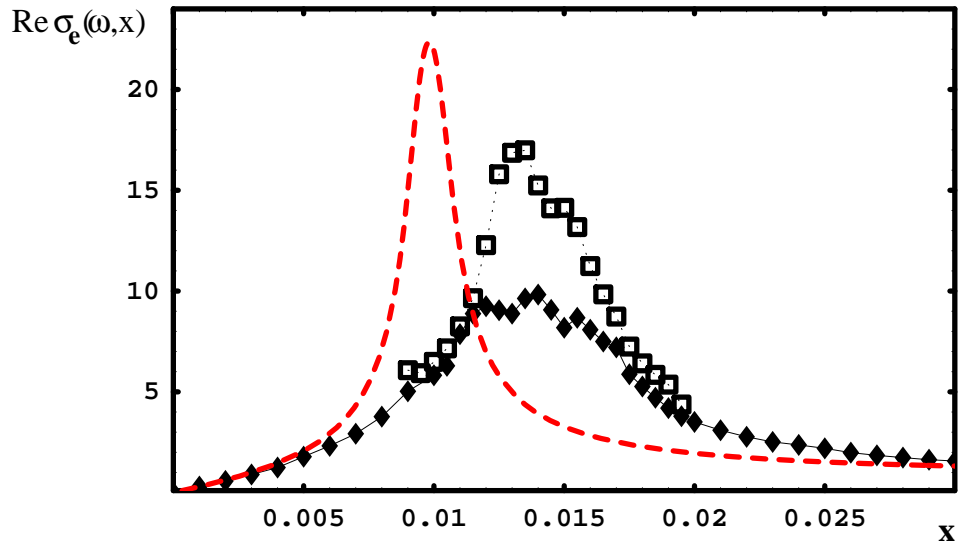, width=5in}
\vskip5.pc
\caption{
Same as Fig.\ 3 but for Re$\sigma_e(\omega, x)/\sigma_N$.  Solid and
dotted lines are guides to the eye.  The RCA calculations
for the larger size are carried out only between x = 0.0095 and x = 0.02.
}
\label{fig4}
\end{figure}
\newpage

\begin{figure}
 \epsfig{file=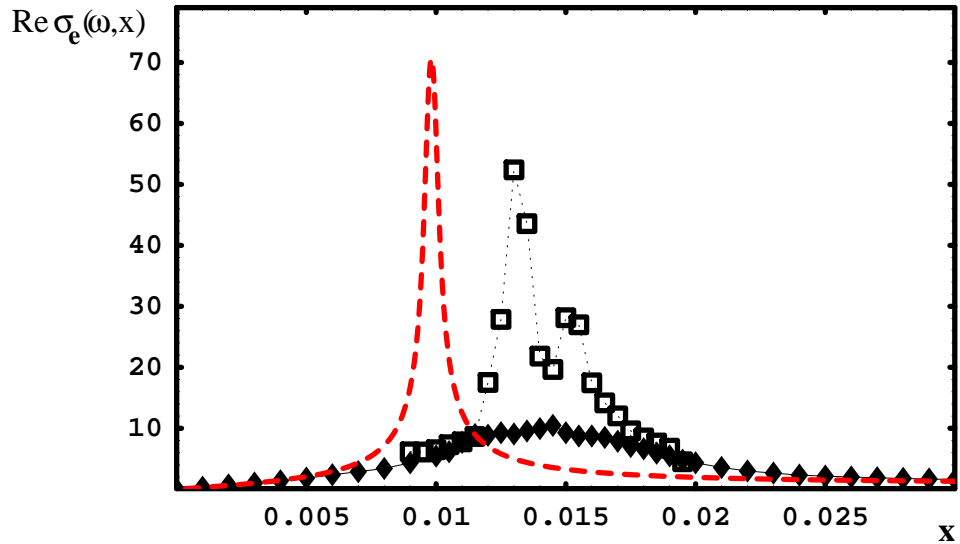, width=5in}
\vskip5.pc
\caption{
Same as Fig.\ 4 but for $\omega = 0.0001\omega_0$,
$\xi_{ab}/a = 4.74$.  In this case, the full diamonds
represent twenty correlated realizations of the RCA for a
$140 \times 140$ lattice, and the open squares are two realizations
for a lattice of size $500 \times
500$.  This latter calculation is carried out only between
$x = 0.01$ and $x = 0.02$.  Light solid and dotted lines 
just connect the calculated points.  Dashed line with no data points: EMA.
}
\label{fig5}
\end{figure}
\newpage

\begin{figure}
 \epsfig{file=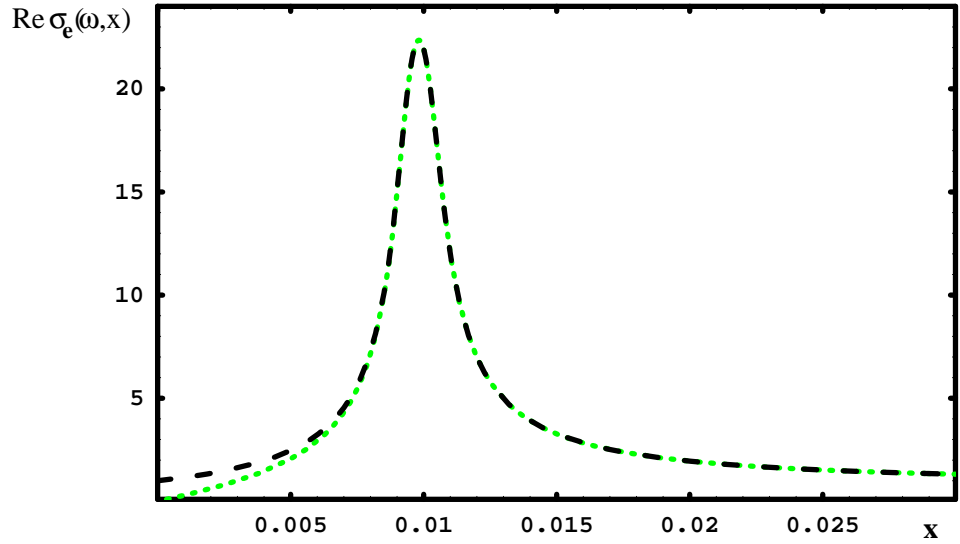, width=5in}
\vskip5.pc
\caption{
Re$\sigma_e(\omega,x)/\sigma_N$ as calculated in the EMA for two cases.
Dotted curve: $\sigma_S(\omega) = iA/\omega$, $\sigma_N = $ const.; 
dashed curve: $\sigma_S(\omega) = iA/\omega + \sigma_N$, $\sigma_N =$ const.
Units are same as in Figs.\ 2-5.
}
\label{fig6}
\end{figure}

% \end{multicols}

\end{document}